# Synchronization with mismatched synaptic delays: A unique role of elastic neuronal latency

Roni Vardi,[1] Reut Timor,[2] Shimon Marom,[3] Moshe Abeles,[1] and Ido Kanter[1,2*]

[1]Gonda Interdisciplinary Brain Research Center, and the Goodman Faculty of Life Sciences, Bar-Ilan University Ramat-Gan 52900, Israel.

[2]Deptartment of Physics, Bar-Ilan University, Ramat-Gan 52900, Israel.

[3]Network Biology Research Laboratories, Technion – Israel Institute of Technology, Haifa 32000, Israel.

**We show that the unavoidable increase in neuronal response latency to ongoing stimulation serves as a nonuniform gradual stretching of neuronal circuit delay loops and emerges as an essential mechanism in the formation of various types of neuronal timers. Synchronization emerges as a transient phenomenon without predefined precise matched synaptic delays. These findings are described in an experimental procedure where conditioned stimulations were enforced on a circuit of neurons embedded within a large-scale network of cortical cells in-vitro, and are corroborated by neuronal simulations. They evidence a new cortical timescale based on tens of $\mu$s stretching of neuronal circuit delay loops per spike, and with realistic delays of a few milliseconds, synchronization emerges for a finite fraction of neuronal circuit delays.**

Psychological and physiological considerations entail that formation and functionality of neuronal cell assemblies [1-3] depend upon synchronized repeated activation such as zero-lag synchronization [4-6]. Several mechanisms for the emergence of this phenomenon have been suggested [7-8], including the global network quantity, the greatest common divisor of neuronal circuit delay loops [9-10] (see the role of the GCD in [25]). However, they require strict biological prerequisites such as precise matched delays and connectivity, and synchronization is represented as a stationary mode of activity instead of a transient phenomenon [11-12].

Similarly, the harmonic activity of interconnected computational and communication devices requires accurate specifications, reliable units and precise wiring. Recently, these demands have been found to apply as well to the emergence of synchronous activity in other networks of threshold elements such as coupled laser networks [13-16]. Viewed from this perspective and given the compromised reliability of their building blocks, the capacity of biological neural networks to generate functional synchronizations on a millisecond time scale is puzzling.

We present in this Letter theoretical evidence corroborated experimentally to show how the apparent variability in brain building blocks can turn into an advantage. We show that the unavoidable increase in neuronal response latency [17-18] to ongoing stimulation serves as a nonuniform gradual stretching of neuronal circuit delay loops. This apparent nuisance is revealed to be an essential mechanism in various types of neuronal timers since the emergence of synchronization emerges as a transient phenomenon and without predefined precise matched synaptic delays. These findings are described in an experimental procedure where conditioned stimulations [10] were enforced on a circuit of neurons embedded within a large-scale network of cortical cells in-vitro [10,19-21], and are corroborated and extended by simulations of circuits composed of Hodgkin-Huxley neurons [22-23] with time-dependent latencies. An exhaustive enumeration of the space of a neuronal circuit with realistic delays of a few milliseconds indicates that synchronization is a common phenomenon that occurs for a finite fraction of neuronal circuit delays. These findings announce a new cortical timescales based on tens of $\mu$s stretching of neuronal circuit delay loops per spike.

At the single neuron level, the most significant feature that appears to work against the formation of millisecond scale synchronies is the tendency of neurons when stimulated repeatedly to gradually change their stimulus-response delay over a few milliseconds [17-18]. To exemplify this neuronal feature, experiments were conducted on cultured cortical neurons that were functionally isolated from their network by a pharmacological block of both glutamatergic and GABAergic synapses [18-19] (see Methods in [25]). Schematic of a neuron with 1:1 responses to a stimulation rate of 10 Hz (Fig. 1(a)) and the experimental results for the neuronal response latency, time-lag between stimulation and evoked spike, are shown (Fig. 1(b)). The results indicate that the latency increases by ~4 ms in ~100 s until critical latency is reached, where the neuron enters the intermittent phase [18]. The average increment of the latency per spike is ~4 $\mu$s, which represents a new, finer time scale of cortical dynamics. The increase in the neuronal latency (internal dynamic) can be equivalently attributed to the extension of the self-feedback delay whereas the neuronal response latency remains unchanged. For instance, the increase of ~2 ms in the neuronal latency after ~400 spikes is equivalent to the scenario where the neuronal response latency is unchanged and the self-feedback delay is extended to 100+2 ms (Fig. 1(c)).

. While the precise underlying mechanisms might be system-specific (types of ionic channels and spatial considerations), it is generally agreed that the increase in latency reflects a decline in the exciting conductances and is fully reversible [17-18].

To analyze the impact of dynamic neuronal response latency at the circuit level, we artificially generated conditioned stimulations of a circuit of neurons embedded within a large scale network of cortical cells *in-vitro* (see Methods in [25]). Our first experimental design consisted of three neurons forming a heterogeneous neuronal circuit (Fig. 2(a)), where the dynamics is initiated by an electrical stimulation to neuron A [10]. Conditioned stimulations were given according to the connectivity of the circuit; e.g. conditioned to evoked spike from neuron B neurons A and C are stimulated after $\tau$ ms. Since the circuit (Fig. 2(a)) consists of $2\tau$ (A→B→A) and $5\tau$ loops (A→B→C→A) and the greatest common divisor GCD(2,5)=1, neuronal activity relaxes to zero-lag synchronization (ZLS) as was theoretically predicted [9,14]. Indeed standard simulations [9-10] of such a circuit indicate that after a short transient of $7\tau$ ms ZLS is achieved (Fig. 2(b)).

A slightly modified circuit is presented in Fig. 2(c) with $2\tau$ and $5\tau+\varepsilon$ loops where ZLS is no longer a solution. Assume that the increase in the neuronal latency per spike, $\Delta$, is independent of the current latency (e.g. a linear fit up to ~1100 spikes in Fig. 1(b)). After q evoked spikes per neuron, the effective delay loops are $2\tau+2q\Delta$ and $5\tau+\varepsilon+3q\Delta$, where the stretching of each unidirectional delay is illustrated by a red bar (Fig. 2(d)). To verify whether the ratio 2:5 between the two loops is restored the equation below is applied

$$\frac{2\tau + 2q\Delta}{2} = \frac{5\tau + \varepsilon + 3q\Delta}{5} \quad (1)$$

which indicates that after

$$q = \frac{\varepsilon}{2\Delta} \quad (2)$$

spikes per neuron ZLS is temporarily restored. Quantitative simulations of such a circuit model composed of Hodgkin-Huxley (HH) neurons with $\Delta=0.004$ ms, $\varepsilon=3$ ms and artificially ignoring a second spike arriving at a neuron in a refractory period of 4 ms indicated thatindeed the time-lag between the evoked spikes of neurons A and B decays linearly to zero after q~3/(2*0.004)=375 (eq. 2) and is followed by desynchronization (Fig. 2(e)). This phenomenon is robust for the case where all spikes are taken into account; however, when doubling the incoming coupling strength to neuron A, ZLS is achieved much faster (Fig. 2(e)). Experimental results with $\varepsilon$=1.5, 2.5 and 3 ms and $\tau$=80 ms indicated that the variable $\varepsilon$ functions as a timer where the transient time to achieve synchronization increases with $\varepsilon$ (Fig. 2(f)-(h)). The noisy quantized behavior is an outcome of the experimental time resolution, 0.5 ms in the timing of stimulations and 1/16 ms in the identification of evoked spikes, as is evidenced by the large fluctuations around the edges of the latency stairs (Fig. 2(f)-(h)).

The timer (Fig. 2) depicts experimentally synchronous activity with coordinated mismatches of ~1-4 ms over synaptic delays of ~80-100 ms, where such slightly relative imprecise delays still represent inflexible prerequisite biological conditions. Furthermore, transient periods to synchronous activity of a few hundreds of $\tau$ ms are also beyond the typical cortical response timescale. Producing much shorter delays which are relevant to cortical dynamics as well as monitoring circuits composed of much larger numbers of neurons are currently beyond our experimental capabilities. Nevertheless, these types of neuronal circuits are scalable, where the stretching of synfire chains [5] (loops) increases linearly with the number of their relays (Fig. S3 in [25]). In addition, neuronal response latencies increase significantly faster (by one order of magnitude) in the initial spiking activity (Figs. S1 and S4 in [25]). Both of these ingredients are expected to significantly shortenthe transient to synchronization.

The quantity at the basis of this timer is the [entire loop latency stretching]/[unit delay $\tau$], which in our example (Fig. 2) is $\Delta$ for the shorter loop (2$\tau$) and is only 3$\Delta$/5 for the longer loop (5$\tau$) (eq. 1). Hence, the relative stretching of the shorter loop is faster and compensates for the redundancy $\varepsilon$ of the longer loop and synchronization is achieved. The inverse scenario, leading to another type of timers, is shown by the heterogeneous circuit with 5 neurons and two loops, 4$\tau$ and 6$\tau$ (Fig. 3(a)). Since GCD(4,6)=2, the neuronal activity relaxes into two neuronal clusters, (A,B,D) and (C,E), where each neuronal cluster is in ZLS. Simulations of such neuronal circuits confirm that after a transient of 9$\tau$, two clusters are formed (Fig. 3(b)). A modified circuit with 4$\tau$ and 6$\tau$-$\varepsilon$ loops does not maintain ZLS between neurons A and B (Fig. 3(c)). After a neuron evoked q spikes, its incoming unidirectional delays elongate by q$\Delta$ as represented by the

red bars (Fig. 3(d)). The restored ratio 4:6 between the two loops is given by the equation

$$\frac{4\tau + 2q\Delta}{4} = \frac{6\tau - \varepsilon + 5q\Delta}{6} \quad (3)$$

which results in

$$q = \frac{\varepsilon}{2\Delta} \quad (4)$$

and the restored duration is $2q\tau$, as the firing period is $GCD^*\tau=2\tau$ (Fig. 3(b)). Simulations with $\Delta=0.004$ ms, $\varepsilon=3$ and ignoring spikes in a refractory period of 4 ms to a neuron indicate that $q\sim 3/(2*0.004)=375$ (Fig. 3(e)), whereas including all spikes in the dynamics results in the same phenomenon but a shorter transient (Fig. 3(e)). Experimental results with $\varepsilon=3$ ms and $\tau=60$ ms indicate that ZLS is achieved after $q\sim 200$ spikes (Fig. 3(f)), and exemplify the robustness of the proposed mechanism to non-identical profiles (mostly concave) of the 5 neuronal response latencies (Fig. S4 in [25]). Actually, the emergence of ZLS is controlled by the relative stretching of the two loops $\Delta L_{4,6}=(L_A+L_B+L_C+L_D+L_E)/6-(L_A+L_B)/4$ (Fig. 3(g)). The prominent feature of this timer is that after a transient trajectory, synchronization appears as a stationary phenomenon (Fig. 3(e)-(f)), unlike the momentary synchronous activity of the first timer (Fig. 2(e)-(h)).

These two types of timers are not an outcome of their different GCD, but rather originated from their intrinsic structures. The first moment where two neuronal clusters emerge (Fig. 3(a)-(b)), neuron A receives two evoked spikes via two identical delay routes; A→B→A and B→C→D→E→A. As the dynamics evolves, the first route becomes relatively shorter since its expansion comprises only two latencies (Fig. 3(d)). Hence, a spike absorbed via the second route falls into the refractory period of the spike that arrived earlier via the first route. Practically, the circuit degenerates into two face-to-face (A↔B) neurons. Similarly, for the first timer when ZLS first emerges (Fig. 2(a)) neuron A receives two evoked spikes via two identical delay routes; B→C→A and A→B→A→B→A. The expansion of the second route is relatively larger, since it comprises more latencies and consequently its spike falls into the refractory period of the first route. Practically the circuit degenerates into a directed heterogeneous loop, A→B→C→A, where synchronous activity cannot be maintained.

A phenomenological approach to characterizing the emergence of neuronal timers requires a mapping between actual circuit delays (Figs. 2(d) and 3(d)) and the dynamics of each neuronal response latency (Fig. 1(b)). The key element is the increase in neuronal response latency with stimulation frequency (Fig. 4(a)), where it can be approximated by a linear fit before arriving at an intermittent phase (Fig. 4(b)). For sufficient high frequencies the increase in the neuronal response latency per spike saturates and converges to $\Delta L_0$ (Fig, 4(b)). The latency now has two main components; an abrupt increase per spike and slow recovery, fading (Fig 4(c)). Although neuronal response latencies are highly varied (Figs. S1 and S4 in [25]), in principle for each delay the equation below can be written

$$d\tau_{ij}/dt = \Delta L_0 \delta(t-t_i) - dF(t-t_i)/dt \quad (5)$$

where $\tau_{ij}$ is the delay from neuron j to i, $t_i$ is the timing of the last spike of neuron i and the last term stands for the derivative of latency fading (Fig. 4(c)) with respect to the elapsed time since the last spike. This set of circuit delay equations together with the HH circuit equations (see Methods in [25]) comprise the entire network dynamical behavior. Since the number of delays is superior to the number of neurons some of these additional equations are redundant, however, an update ofeach neuronal latency by its internal parameters requires an additional physiological insight [24].

Delays of a few ms are still beyond our experimental capabilities. Nevertheless, we theoretically analyzeda six neuron circuit consisting of X and Y delay loops (X>Y) with a maximal 5 ms latency increase (Fig. 4(d)). The condition to achieve ZLS (in non-trivial locations) is given by

$$\frac{X + 3\Delta}{m} = \frac{Y + 5\Delta}{n}$$

where $\Delta$ represents the current latency increase, n>m and GCD(m,n)=1. One can verify that the larger loop is bounded by

$$Y_1 = \frac{nx}{m} \quad and \quad Y_2 = \frac{nx}{m} + 5\frac{(3n - 5m)}{m}$$

and X>5(5m-3n)/n. For 5≤X≤10 ms and 5≤Y≤20 ms, ZLS is achieved in at least 50% of the possible initial X and Y delays (Fig. 4(e)).

Synchronous flashes and epochs are expected to be a common phenomenon in neuronal brain activity. Their probability of occurrence is anticipated to be enhanced by much shorter biological delays and by long synfire chains along the circuit loops which enhance the stretchability of the neuronal circuit. In a set of simulation studies at the population dynamics (cell assembly) level we demonstrated that the experimentally exemplified new cortical mechanism remains valid (Fig. S6 in [25]), and the circuit activity becomes less sensitive to background fluctuations [9]. This feature is especially crucial to the realization of shorter neuronal loops, where the activity of a single neuron will stop spontaneously due to the relative neuronal refractory period or synaptic fatigue. It is also evident that the variety of possible timers is much larger, in that sub-neuronal circuits might consecutively pass through a few types of synchronous activities during the entire latency increase and also in the intermittent phase where the neuronal response latency fluctuates by few ms [18].

The authors thank Vladimir and Elleonora Lyakhov for invaluable technical assistance. The research has received funding from the European Union Seventh Framework Program FP7 under grant agreement 269459 to SM and grant of the Ministry of Science and Technology of the State of Israel and MATERA grant agreement 3-7878 to SM.

**\*Corresponding author**IK (ido.kanter@biu.ac.il )

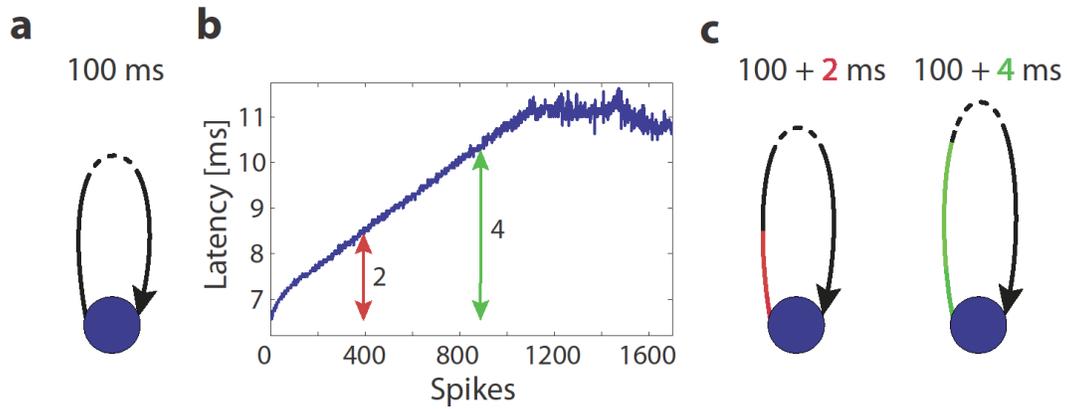

FIG. 1 (color online). (a) Schematic of a neuron with a stimulation of 10 Hz. (b) Experimental results for a neuronal response latency of 1(a) calculated at a time resolution of 1/16 ms (The type of stimulations is given in the Method section in [25]). The latency increases by ~4 ms in ~100 s (on the average, an increase of ~4 μs per spike), until the neuron enters the intermittent phase. The serial spike number where latency increases by ~2ms (red) and ~4 ms (green) is labeled. (c) Schematic showing how the increase in the neuronal response latency shown in 1(b) can also be attributed to the extension of the self-feedback delay while the neuronal response latency remains unchanged.

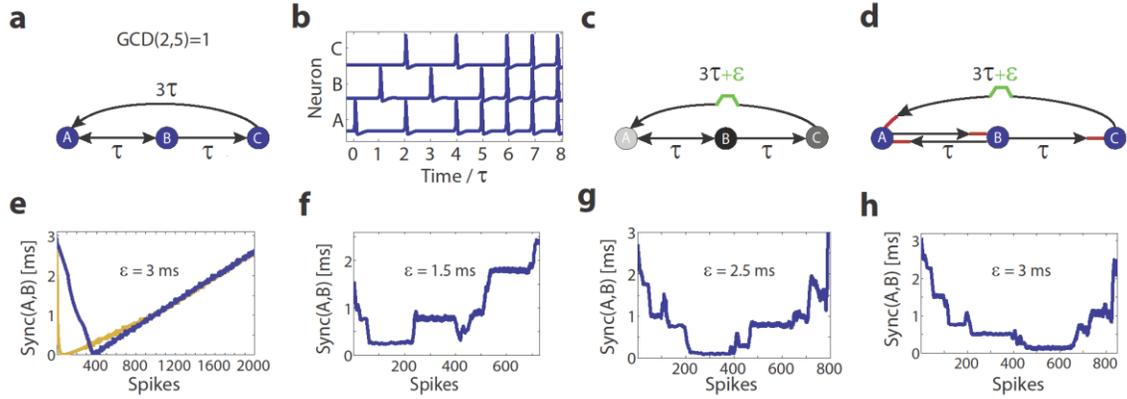

FIG. 2 (color online). (a) Schematic of a heterogeneous neuronal circuit consisting of three neurons with two delay loops, $2\tau$ and $5\tau$, and ZLS as GCD(2,5)=1. Neurons that fire together in the synchronous activity are represented by the same color. (b) Simulations of neuronal circuit 2(a) composed of HH neurons with $\tau$=30 ms. The dynamics are initiated by a stimulation to neuron A. ZLS emerges after a transient of $7\tau$ms. (c) Schematic of the neuronal circuit 2(a), but the $3\tau$ delay from neuron C to A is extended by $\varepsilon$. Neurons are colored differently since ZLS is not a solution. (d) Schematic of the effective extension of each delay, illustrated by the appended red bar. ZLS is temporarily restored, represented by the same color for the three neurons. (e) The time-lag between the evoked spikes of neuron A and B is presented as a function of the number of spike pairs obtained from simulations of circuit 2(d) with $\varepsilon$=3 ms. After a neuron evoked spike, similar to 1(b), its incoming delays are stretched by $\Delta$=0.004 ms. Artificially ignoring spikes arriving at a neuron in a refractory period of 4 ms, ZLS is restored after ~400 spike pairs in accordance with eq. 2 (blue), and when all spikes are taken into account (yellow). (f-h) Experimental results of the time-lag between neurons A and B as a function of spikes for neuronal circuit 2(d) initiated by an electrical stimulation to neuron A with $\varepsilon$=1.5, 2.5 and 3 ms and $\tau$=80 ms. Results are averaged over sliding windows of 10 spikes. To overcome experimental limitation an equivalent circuit was implemented (Fig. S2 in [25]).

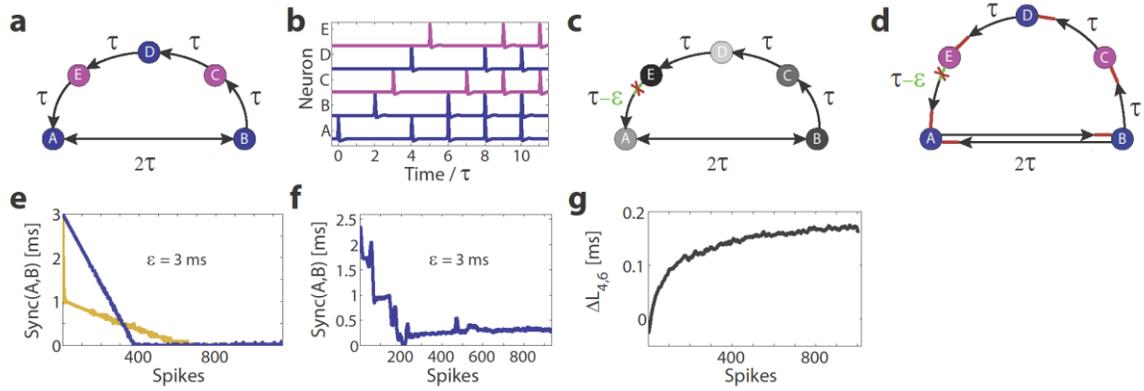

FIG. 3 (color online). (a) Schematic of a heterogeneous neuronal circuit consisting of five neurons with $4\tau$ and $6\tau$ delay loops, and GCD(4,6)=2 neuronal clusters, each in ZLS and represented by the same color. (b) Simulations of circuit 3(a) composed of HH neurons with $\tau=50$ ms. The dynamics is initiated by a stimulation to neuron A. The synchronous activity of the 2-cluster emerges after a transient of $9\tau$ms. (c) Schematic of circuit 3(a), but the $\tau$ delay from neuron E to A is shortened by $\varepsilon$ (represented by the red x). Neurons are not synchronized and colored differently. (d) Schematic of the effective stretching is illustrated by the appended red bar for each directed delay. (e) The time-lag between neuron A and B is presented as a function of spikes for simulations of circuit 2(d), where after a neuron evokes a spike its incoming delays are stretched by $\Delta=0.004$ ms and $\varepsilon=3$ms. Ignoring spikes arriving at a neuron in a refractory period of 4 ms (blue) and taking into account all spikes in the dynamics (yellow). (f) Experimental results of the time-lag between neurons A and B as a function of spikes for neuronal circuit 3(d) initiated by an electrical stimulation to neuron A with $\varepsilon=3$ ms and $\tau=60$ ms as in 2(f). To overcome experimental limitation an equivalent circuit was implemented (Fig. S5 in [25]). (g) Experimental results of the difference between the relative stretching of the two circuit loops $\Delta L_{4,6}=(L_A+L_B+L_C+L_D+L_E)/6-(L_A+L_B)/4$ as a function of spikes (eq. 3).

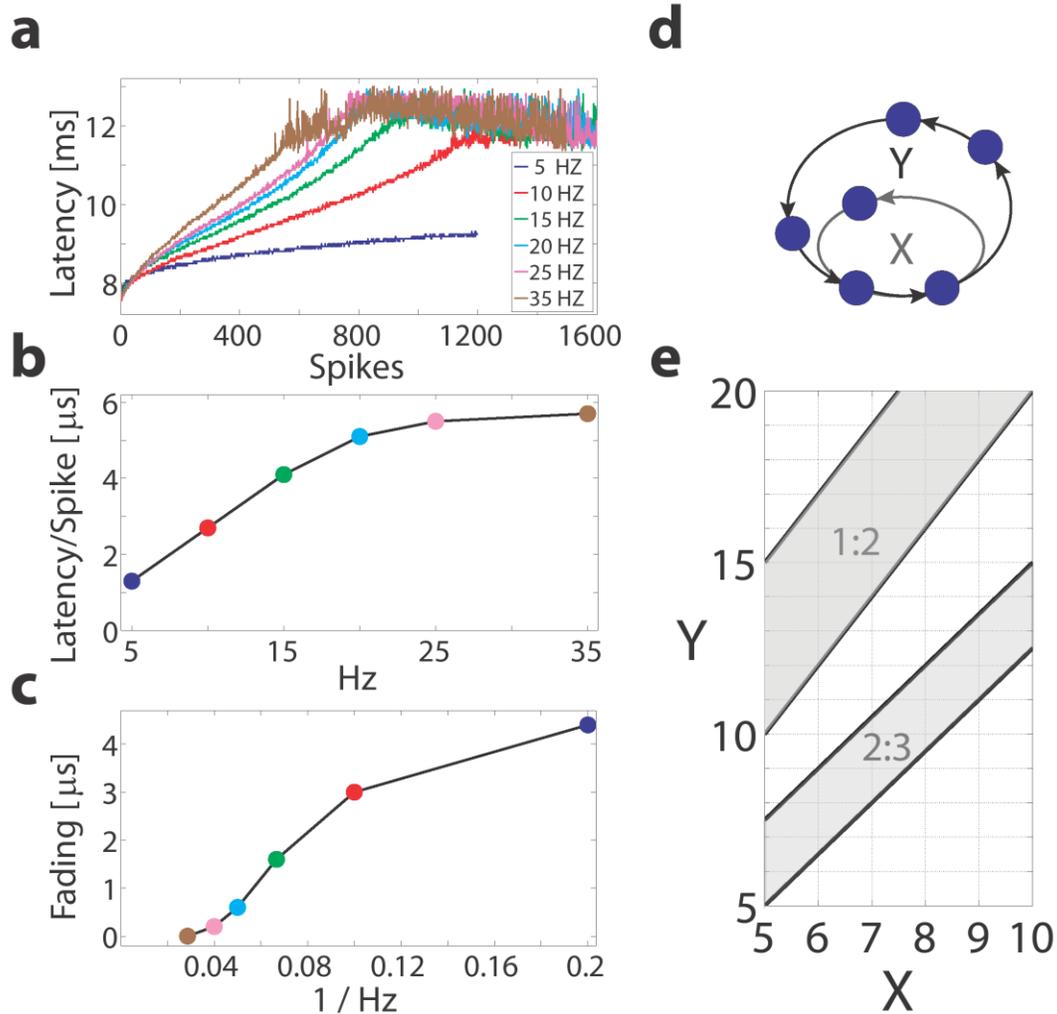

FIG. 4 (color online). (a) Neuronal response latency as a function of spikes for various stimulation frequencies in the range of 5-35 Hz. (b) The average increase in the neuronal response latency per spike, before arriving at an intermittent phase, is obtained using a linear fit for each stimulation frequency in 4(a). (c) Fading in the neuronal response latency as a function of the elapsed time since the last spike. Results are estimated from 4(b) where 35 Hz is taken to be the asymptotic value for high stimulation frequencies. (d) Schematic of a heterogeneous neuronal circuit consisting of two delay loops, X and Y (Y>X), where delays among the six neurons are chosen arbitrarily. (e) Initial conditions for 5≤X≤10 ms and 5≤Y≤20 ms leading to ZLS under the assumption of a maximal 5 ms increase in the response latency of each neuron. Initial (X,Y) delay loops leading to delay loops obeying the ratios 1:2 or 2:3 are marked by the two bounded gray regions. The fraction of initial (X,Y) leading to ZLS is 50%.

# Supplementary Material

# Synchronization with mismatched synaptic delays: A unique role of elastic neuronal latency

Roni Vardi[1], Reut Timor[2], Shimon Marom[3], Moshe Abeles,[1] and Ido Kanter[1,2]

[1]*Gonda Interdisciplinary Brain Research Center, and the Goodman Faculty of Life Sciences, Bar-Ilan University Ramat-Gan 52900, Israel.*

[2]*Deptartment of Physics, Bar-Ilan University, Ramat-Gan 52900, Israel.*

[3]*Network Biology Research Laboratories, Technion – Israel Institute of Technology, Haifa 32000, Israel.*

## 1. The role of the GCD

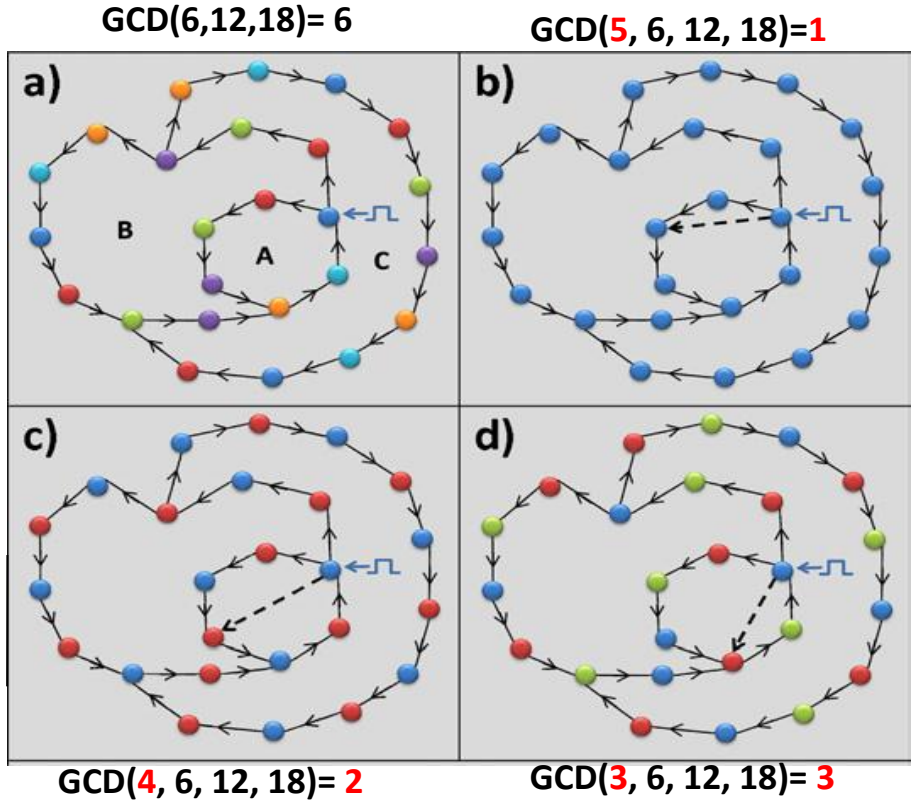

The nonlocal role of the GCD is exemplified by the above figure where a circuit consists of 25 nodes and 3 loops with total delays of $6\tau, 12\tau$ and $18\tau$ (boundaries of areas A, A+B and C, respectively, where $\tau$ is a unit delay between two connected nodes), and a stimulus to one node (for more details see reference [9] of the manuscript). Nodes split into 6-clusters following the GCD(6,12,18)= 6. (b) With an additional unidirectional connection (dashed arrow) and a loop of $5\tau$, the GCD(5,6,12,18)=1 and the circuit is in ZLS. (c) An additional loop of $4\tau$ as in (b), where GCD(4,6,12,18)= 2 and the circuit is in 2-clusters. (d) An additional loop of $3\tau$ as in (b), where GCD(3,6,12,18)= 3 and the circuit is in 3-clusters.

## 2. Supplementary Methods:

### 2.1 Parameters used for the Hodgkin Huxley Model

To explore the local and nonlocal behavior of cortical neural activity, a realistic numerical model was designed in which the dynamic behavior of a solitary neuron, delays and connectivity strength were taken into account [1,2]. Each neural cell was simulated using the well- known Hodgkin Huxley model [3] (HH).

For every neuron (i) in the network the membrane potential $V^i$ is described by the following differential equation:

(1)

$$c_m \frac{dV^i}{dt} = -g_{Na} m^{i^3} h^i (V^i - E_{Na}) - g_k n^{i^4} (V^i - E_k) - g_L (V^i - E_L) + I^i_{syn} + I^i_{ext} + I^i_{noise}$$

where $C_m$=1 µF/cm² is the membrane capacitance. The constants $g_{Na}$=120mS/cm², $g_k$=36mS/cm² and $g_L$=0.3mS/cm² are the maximal conductances of their corresponding channels, and $E_{Na}$=50mV, $E_k$=-77mV and $E_L$=-54.5mV are the corresponding reversal potentials. The voltage-gated ion channels m, n and h represent the activation and inactivation of the sodium and potassium channels and can be described by the following three differential equations:

(2)

$$\frac{dX^i}{dt} = \alpha^i_X(V^i)(1 - X^i) - \beta^i_X(V^i) X^i$$

Where X = m, n, h. The experimentally fitted voltage-dependent transition rates are:

(3)

$$\alpha_m(V) = \frac{0.1(V+40)}{1 - \exp(-0.1(V+40))}$$
$$\beta_m(V) = 4\exp(-V+65/18)$$
$$\alpha_n(V) = \frac{0.01(V+55)}{1 - \exp(-0.1(V+55))}$$
$$\beta_n(V) = 0.125\exp(-V+65/80)$$
$$\alpha_h(V) = 0.07\exp(-V+65/20)$$
$$\beta_h(V) = \frac{1}{1 + \exp(-0.1(V+35))}$$

In the absence of any type of noise or synaptic influence on the neuron, the steady state $V^i_{rest}$=-65mV is stable for $I_{ext}$<9.78 µA/cm². However when 9.78<$I_{ext}$<154.5 µA/cm² the HH neuron starts to fire periodically [4]. Our simulations were adjusted to imitate the behavior of a random biological neural cell.

The synaptic transmission between neurons is modeled by a postsynaptic conductance change with the form of an α function:

(4)

$$\alpha(t) = \frac{\exp(-t/\tau_d) - \exp(-t/\tau_r)}{\tau_d - \tau_r}$$

where the parameters $\tau_d$=10 ms and $\tau_r$=1 ms stand for the decay and rise time of the function and determine the duration of the response. The synaptic current $I^i_{syn}(t)$ takes the form:

(5)

$$I^i{}_{syn}(t) = -g_{max} \sum_j \sum_{t_j^{sp}} \alpha(t - t_j^{sp} - \delta_{ij})(V - E_{syn})$$

Here, {j} is the group of neurons coupled to neuron i. The internal sum is taken over the train of pre-synaptic spikes occurring at $t_j^{sp}$ of a neuron j in the group. Excitatory and inhibitory transmissions were differentiated by setting the synaptic reversal potential to be $E_{syn}$ = 0 mV or $E_{syn}$ = -80 mV, respectively. gmax=1.2 ms/cm$^2$ describes the maximal synaptic conductance between neurons i and j.

We integrated the set of differential equations numerically using Heun's method. The time step of the integration was 0.002 ms.

### 2.2 Population dynamics- The Hodgkin Huxley Model

In our network model, each node represents one cortical patch comprised of N neurons. Every neural cell was simulated using the well known HH model [3] and in terms of biological properties it was assumed that distant cortico-cortical connections are (almost) exclusively excitatory whereas local connections are both excitatory and inhibitory [5,6]. In this framework, cortical areas are connected reciprocally across the two hemispheres and within a single hemisphere [7,8], where small functional cortical units (patches) connect to other cortical patches in a pseudo random manner. The number of patches to which a single patch connects varies considerably, where typically, it grows as the square root of the number of cortical neurons [9] resulting for a mouse in 3 to 6 [6], and most likely for humans in roughly 150.

A connection between neurons belonging to different nodes was excitatory only and was selected with a probability of $p_{out}$=0.2. The delay $\tau_{ij}$ between a pair of neurons belonging to different nodes i and j, represents the time it takes an action potential to travel through the axon from the pre synaptic neuron i to the post synaptic neuron j and was taken from a uniform distribution in the range [τ−1, τ+1] ms, where τ was selected to be 30 ms.

In order to simplify our model, populations were composed of excitatory neurons only. We integrated the set of differential equations numerically using Heun's method. The time step of the integration was 0.02 ms.

**2.3 Methods**

*Cell preparation*

Cortical neurons were obtained from newborn rats (Sprague-Dawley) within 48 h after birth using mechanical and enzymatic procedures described in earlier studies [2,10-14]. Rats were euthanized by CO2 treatment according to protocols approved by the National Institutes of Health. The neurons were plated directly onto substrate-integrated multi-electrode arrays and allowed to develop functionally and structurally mature networks over a time period of 2–3 weeks. The number of plated neurons in a typical network is on the order of 1,300,000, covering an area of about 380 mm$^2$. The preparations were bathed in MEM supplemented with heat-inactivated horse serum (5%), glutamine (0.5 mM), glucose (20 mM), and gentamicin (10 g/ml), and maintained in an atmosphere of 37 ◦C, 5% CO2, and 95% air in an incubator as well as during the recording phases.

All experiments were conducted in the standard growth medium, supplemented with 25 µM Bicuculline, 50.26 µM CNQX (6-cyano-7-nitroquinoxaline-2,3-dione) and 401.6 µM APV (amino-5-phosphonovaleric acid); this cocktail of synaptic blockers made the spontaneous network activity sparse. At least an hour was allowed for stabilization of the effect.

*Measurements and stimulation*

An array of 60 Ti/Au/TiN extracellular electrodes, 30 µm in diameter, and spaced either 500 or 200 µm from each other (Multi-ChannelSystems, Reutlingen, Germany) were used. The insulation layer (silicon nitride) was pre-treated with polyethyleneimine (Sigma, 0.01% in 0.1M Borate buffer solution). A commercial amplifier (MEA-1060-inv-BC,MCS, Reutlingen, Germany) with frequency limits of 150–3000 Hz and a gain of ×1024 was used. Mono-phasic square voltage pulses (100-500 µs, 100–900 mV) were applied through extracellular electrodes. The data were digitized using data acquisition board (PD2-MF-64-3M/12H,UEI, Walpole, MA, USA). Each channel was sampled at a frequency of 16K sample/s. Action potentials were detected on-line by threshold crossing. Data processing and conditioned stimulation were performed using Simulink (The Mathworks, Natick, MA, USA) based xPC target application.

*Cell selection*

Each circuit node was represented by a stimulation source (source electrode) and a target for the stimulation – the recorded electrode (target electrode). The stimulation electrodes (source and target) were selected as the ones that evoked well-isolated and well-formed spikes and a reliable response with a high signal-to-noise ratio. This examination was done with a stimulus intensity of 800 mV, after 30 repetitions at a frequency of 5Hz.

*Stimulation control*

The activity of all target electrodes for each stimulation was collected and entailed stimuli were delivered in accordance to the adjacency matrix.

FIG. S1: Neuronal response latencies – first time-controller

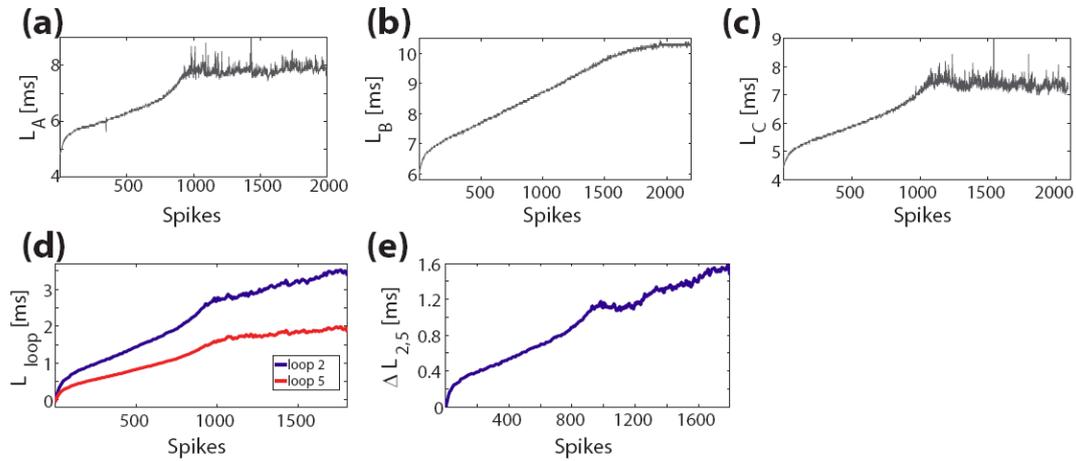

FIG. S1: (a)-(c) Neuronal response latencies ($L_A$, $L_B$, $L_C$) of the 3 neurons used in the experiment Fig. 2(f)-(h) of the manuscript. (d) The experimental elongation of the two circuit loops per unit delay, $\tau$, as a function of spikes: $\Delta L_2=(L_A+L_B)/2$ and $\Delta L_5=(L_A+L_B+L_C)/5$. (e) The difference between the relative elongation of the two circuit loops $\Delta L_{2,5}=(L_A+L_B+L_C)/5-(L_A+L_B)/2$ as a function of spikes.

Fig. S2: Shifted experimental neuronal circuit for the first time-controller

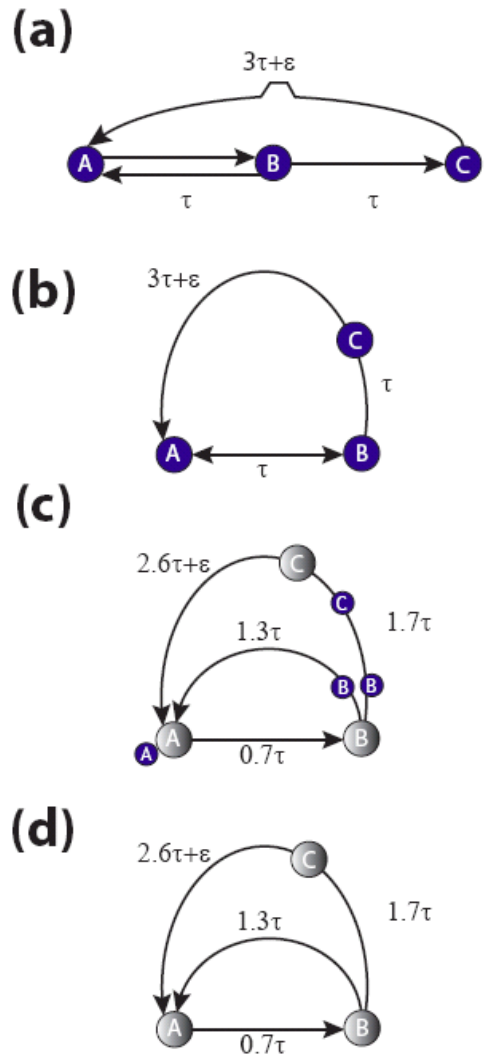

FIG S2: (a) Schematic of heterogeneous neuronal circuit 2(a) in the manuscript. The circuit consists of three neurons with two delay loops, $2\tau$ and $5\tau+\varepsilon$, and ZLS as GCD(2,5)=1 with $\varepsilon=0$. (b) An equivalent schematic of circuit 2(a). (c) The circuit that was implemented to overcome the experimental limitation of the lack of independent simultaneous stimulators. For the sake of clarity, the original nodes of circuit 2(a)-(b) before shifting are presented as small blue nodes in their original location. The three large gray nodes represent the actual location of the neurons in the experimental circuit. Blue node B is plotted twice since it drives nodes A and C. Gray node B fires $0.3\tau$ before blue node B. Gray node C fires $0.4\tau$ after blue node C. The timings of the spikes measured in the experiment on the gray neurons are shifted to the originalblue neurons. (d) Schematic of the shifted experimental circuit.

FIG. S3: Enhanced stretchability by longer synfire chains

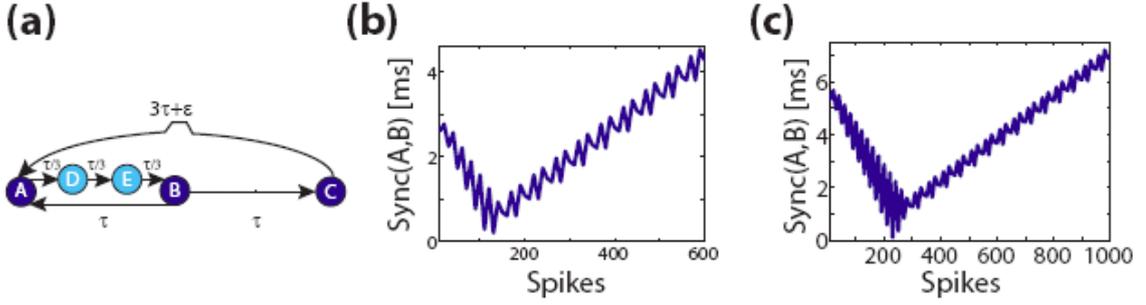

FIG S3: (a) Schematic of heterogeneous neuronal circuit 2(a) in the manuscript with 2 additional neurons (light blue) on the route from A to B. The circuit consists of five neurons while the total delay loops remain unchanged, $2\tau$ and $5\tau+\varepsilon$, and ZLS is reached as GCD(2,5)=1 with $\varepsilon=0$. (b)-(c) The time-lag between the evoked spikes of neuron A and B is presented as a function of the counted spike pairs obtained from simulations of circuit 3(a) composed of HH neurons with $\varepsilon=3$ ms (b) and 6 ms (c). After a neuron evoked spike, similar to fig. 2(e) in the manuscript its incoming delays are stretched by $\Delta=0.004$ ms (4 µs). Artificially ignoring spikes arriving at a neuron in a refractory period of 4 ms, ZLS is temporarily restored after ~130 (b) and ~250 (c) spike pairs. These transients are in accordance with q=150 (b), q=300 (c) obtained from the equation:

$$\frac{2\tau + 4q\Delta}{2} = \frac{5\tau + \varepsilon + 5q\Delta}{5} \quad (1)$$

results in

$$q = \frac{\varepsilon}{5\Delta} \quad (2)$$

Without neurons D and E, Fig. 2(a) in the manuscript, ZLS is achieved for q~400, since q=$\varepsilon$/2$\Delta$ only. The addition of two neurons improves the stretchability of the neuronal circuit loops, however, the relative elongation of the shorter loop ($2\tau$) is enhanced further. This feature leads to a shorter transient time to achieve temporary ZLS.

FIG S4: Neuronal response latencies – second time-controller

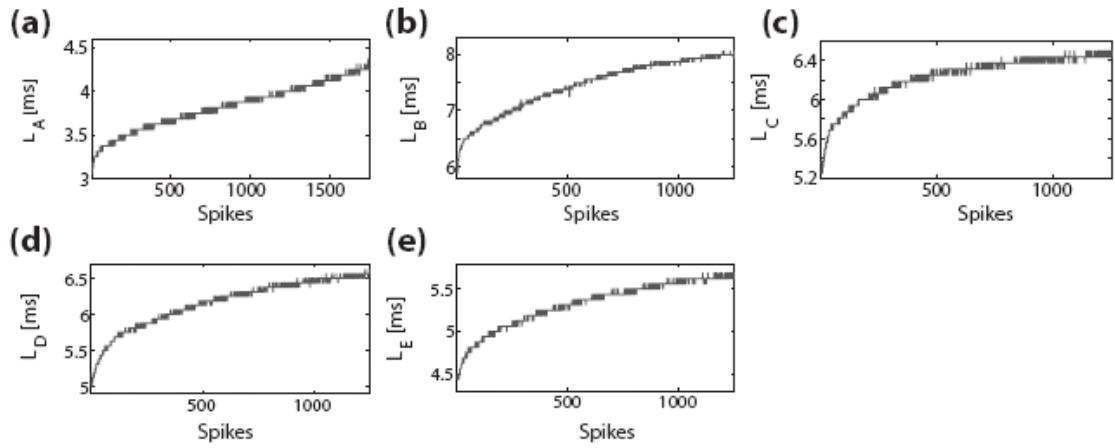

FIG S4: (a)-(e), Neuronal response latencies ($L_A$, $L_B$, $L_C$, $L_D$, $L_E$) of the 5 neurons used in the experiment Fig. 3(f)-(g) in the manuscript.

FIG S5: Shifted experimental neuronal circuit for the second time-controller

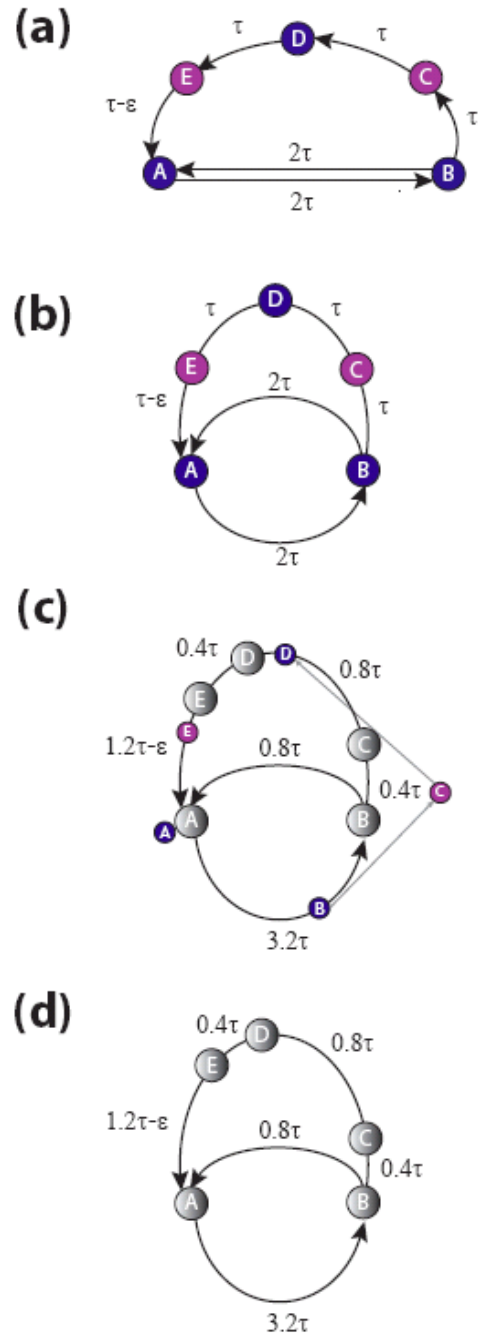

FIG S5: (a) Schematic of heterogeneous neuronal circuit 3(a) in the manuscript. The circuit consists of five neurons with $4\tau$ and $6\tau-\varepsilon$ delay loops, and with $\varepsilon=0$ GCD(4,6)=2 neuronal clusters. (b) An equivalent schematic of circuit S5(a). (c) The circuit that was implemented to overcome the experimental limitation of the lack of independent simultaneous stimulators. For the sake of clarity, the original nodes of circuit S5(a)-(b) before shifting are presented as small blue nodes in their original location. The five large gray nodes represent the actual location of the neurons in the experimental circuit. Gray node B fires $1.2\tau$ after blue node B. Gray node C fires $0.6\tau$ after blue node C. Gray node D fires $0.4\tau$ after blue node D. Gray node E fires $0.2\tau$ before blue node E. The timings of the spikes measured in the experiment on the gray neurons are shifted to the original blue neurons. (d) Schematic of the shifted experimental circuit.

FIG. S6: Population dynamics- The Hodgkin Huxley Model

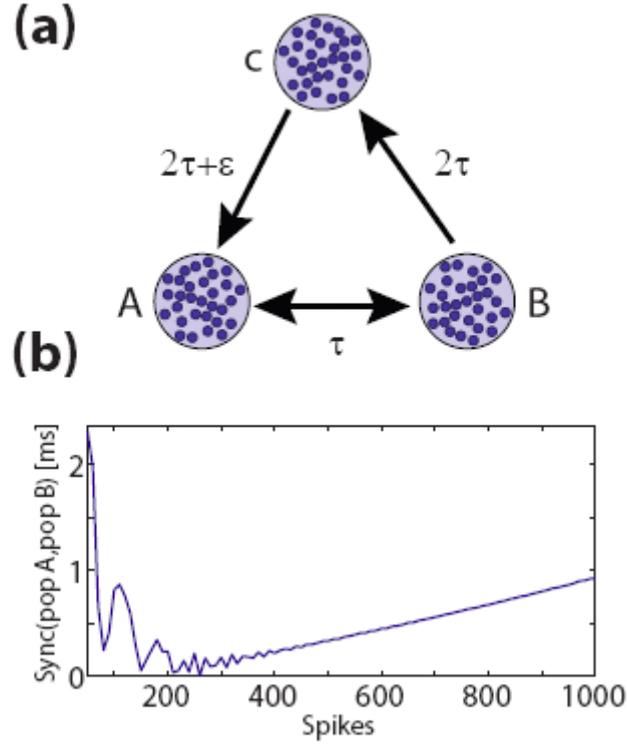

FIG S6: (a) Schematic of a heterogeneous neuronal circuit consisting of three nodes with two delay loops, $2\tau$ and $5\tau+\varepsilon$, and with $\varepsilon=0$ ZLS is achieved as GCD(2,5)=1 (as in Fig. 2(a) of the manuscript). Each one of the three nodes is comprised of 30 Hodgkin Huxley neurons. A connection between neurons belonging to two connecting nodes is excitatory only and selected with a probability of $p_{out}=0.2$, and the delay between them was taken from a uniform distribution in the range $[\tau-1, \tau+1]$ ms, where $\tau$ was selected to be 30 ms (see Supplementary Methods). (b) The time-lag between evoked spikes of population A and B is presented as a function of counted population spike pairs (a population spike is calculated as the average over the spike timing of all neurons comprising the population) obtained from HH simulations of circuit S6(a) with $\varepsilon=3$ ms. After a neuron evoked spike, its incoming delays are stretched by $\Delta=0.004$ ms. ZLS is temporarily restored after ~200 spikes.